\documentclass[prd,aps,floats,tabularx,preprintnumbers,twocolumn]{revtex4}
\usepackage{graphicx, epsfig}
 
\newcommand{\be}{\begin{equation}}
\newcommand{\ee}{\end{equation}}
\newcommand{\bea}{\begin{eqnarray}}
\newcommand{\eea}{\end{eqnarray}}

\begin{document}
 
\preprint{SU-GP-02/11-1}
\preprint{SU-4252-771}
 
\setlength{\unitlength}{1mm}
 
\title{The State of the Dark Energy Equation of State}
 
\author{Alessandro Melchiorri$^\flat$, Laura Mersini$^*$,
Carolina J. \"Odman$^\sharp$ and Mark Trodden$^*$}
\affiliation{
$^\flat$ Astrophysics, Denys Wilkinson Building, University of Oxford, Keble
road, OX1 3RH, Oxford, UK\\
$^*$ Department of Physics, Syracuse University, Syracuse, NY 13244-1130,
USA.\\
$^\sharp$ Astrophysics Group, Cavendish Laboratory, Cambridge University,
Cambridge, U.K.
}
\begin{abstract}
By combining data from seven cosmic microwave background experiments
(including the latest WMAP results)
with large scale structure data, the Hubble parameter measurement
from the Hubble Space Telescope and luminosity measurements of
Type Ia supernovae we demonstrate the bounds on the dark energy equation
of state $w_Q$ to be $-1.38< w_Q <-0.82$ at the $95 \%$ 
confidence level. Although our limit on $w_Q$ is improved with 
respect to previous analyses, cosmological data does not rule out 
the possibility that the equation of state parameter $w_Q$ of the dark 
energy $Q$ is less than -1.
We present a tracking model that ensures $w _Q \le -1$ at recent times
and discuss the observational consequences.
\end{abstract}
 
 
\maketitle
 
\section{Introduction.}
There is now a growing body of evidence that the evolution of the universe 
may be dominated by a dark energy component $Q$, with present-day energy 
density fraction $\Omega_Q \simeq 2/3$~\cite{super1}. 
Although a true cosmological constant
$\Lambda$ may be responsible for the data, it
is also possible that a dynamical mechanism is at work.
One candidate to explain the observations is a
slowly-rolling dynamical scalar ``quintessence''
field~\cite{Wetterich:fm}-\cite{Caldwell:1997ii}.
Another possibility, known as
``k-essence''~\cite{Armendariz-Picon:1999rj}-\cite{Chiba:1999ka}, is
a scalar field with non-canonical kinetic terms in the lagrangian. Dynamical
dark energy
models such as these, and others~\cite{parker}-\cite{bastero-mersini}, have an
equation of state,
$w_{Q}\equiv p_Q/\rho_Q$ which varies with time compared to that of a
cosmological constant
which remains fixed at
$w_{Q=\Lambda}=-1$. Thus, observationally distinguishing a time variation in
the equation of state or finding $w_Q$ different from $-1$ will
rule out a pure cosmological constant as an explanation for the data,
but be consistent with a dynamical solution.
 
In the past years many analyses of several cosmological datasets
 have been produced in order to constrain $w_Q$
(see e.g. \cite{rachel} and references therein).
In these analyses the case of a constant-with-redshift
$w_Q$, in the range $w_Q\geq -1$ was considered.
The assumption of a constant $w_Q$ is based on
several considerations: first of all, since
both the luminosities and angular distances (that are
the fundamental cosmological observables)
depend on $w_Q$ through multiple
integrals, they are not particularly sensitive to
variations of $w_Q$ with redshift (see e.g. \cite{maor1}, \cite{rachel}). 
Therefore, with current data, no strong constraints can be placed on the
redshift-dependence of $w_Q$.
Second, for most of the dynamical models on the market,
the assumption of a piecewise-constant equation of state is
a good approximation for an unbiased determination
of the effective equation of state (\cite{wang0})
\begin{equation}
\label{averageeos}
w_{\rm eff} \sim \frac{\int w_Q(a) \Omega_Q(a) da}{\int \Omega_Q(a) da}
\end{equation}
predicted by the model.
Hence, if the present data is compatible with a constant
$w_Q=-1$, it may not be possible to discriminate between a cosmological
constant
and a dynamical dark energy model.
 
The limitation to $w_Q>-1$, on the contrary,
is a theoretical consideration motivated, for example, by
imposing on matter (for positive energy densities) the null energy condition,
which
states that
$T_{\mu\nu}N^{\mu}N^{\nu}>0$ for all null 4-vectors $N^{\mu}$.
Such energy conditions are often
demanded in order to ensure stability of the theory.
However, theoretical attempts to obtain $w_Q <-1$ have been
considered~\cite{Caldwell:1999ew,Schulz:2001yx,parker,frampton,Ahmed:2002mj} while a careful
analysis of their potential instabilities has been performed in~\cite{CHT}.
 
Moreover, Maor et al.~\cite{maor2} have recently shown that one may construct
a model with a specific z-dependent $w_Q(z)\geq -1$, in which the assumption of
constant $w_Q$ in the analysis can lead to an estimated value $w_{\rm eff}
<-1$.
This further illustrates the necessity of extending dark energy analyses to
values of $w_Q<-1$.

In this paper we combine constraints from a variety of observational data
to determine the currently allowed range of values for the dark
energy equation of state parameter $w_Q$.
The data used here comes from six recent Cosmic Microwave Background
(CMB) experiments, from the power spectrum  of large scale structure
(LSS) in the 2dF 100k galaxy redshift survey, from luminosity measurements
of Type Ia supernovae (SN-Ia)~\cite{super1} and from the Hubble
Space Telescope (HST) measurements of the Hubble parameter.
 
Our analysis method and our datasets are very similar to the one
used in a recent work by Hannestad and Mortsell (\cite{mortsell}).
We will compare our results with those derived in this
earlier paper in the conclusions.

In the next section, we demonstrate the plausibility of $w_Q <-1$ by presenting
a class of theoretical models in which this result may be obtained explicitly. 
In our model the equation of state parameter is approximately piecewise 
constant and hence provides a specific
example of a model which would fit the data described in the remainder of the
paper.
The methods used to obtain combined constraints
on the dark energy equation of state are described in section~\ref{method}.
Our likelihood analysis is presented in section~\ref{results}
and our summary and conclusions are given in section~\ref{conclusions}.
 
\section{A Model with $w_Q < -1$.}
\label{model}
It is a simple exercise to show that a conventional scalar field lagrangian
density cannot yield an equation of state parameter $w_Q <-1$. There are,
however, a number of ways in which the lagrangian can be modified to make $w_Q
<-1$ possible. For example, one may reverse the sign of the kinetic terms,
leading to interesting cosmological and particle physics
behavior~\cite{Caldwell:1999ew}-\cite{CHT}.
 
Let us motivate the study of $w_Q <-1$ cosmologies by describing a class of
models, {\it with positive kinetic energy}, in which such evolution arises. The
model we consider is very much in the spirit of
{\it k-essence}~\cite{Armendariz-Picon:2000dh,Armendariz-Picon:2000ah}.
Comments on the similarities and differences between the two are briefly
discussed at the end of this section.
 
Consider a theory of a real scalar field $\phi$, assumed to be homogeneous,
with a non-canonical kinetic energy term. The lagrangian density is
\begin{equation}
\label{lagrangian}
{\cal L}=f(\phi)g(X)-V(\phi) \ ,
\end{equation}
where $f(\phi)$, $g(X)$ are positive semi-definite functions, $V(\phi)$ is a
potential and $X\equiv {\dot \phi}^2/2$. The energy-momentum tensor for this
field is straightforward to calculate and yields the usual perfect fluid form
with pressure
$p$ and energy density $\rho$ given by
\begin{equation}
\label{pressure}
p={\cal L}=f(\phi)g(X)-V(\phi) \ ,
\end{equation}
\begin{equation}
\label{energydensity}
\rho=\left[2X\frac{dg(X)}{dX}-g(X)\right]f(\phi)+V(\phi) \ .
\end{equation}
Thus, defining $w_{\phi}\equiv p/\rho$ one obtains
\begin{equation}
\label{w}
w_{\phi}=\frac{g(X)f(\phi)-V(\phi)}{\left[2X\frac{dg(X)}{dX}-g(X)\right]f(\phi)+
V(\phi)} \ .
\end{equation}
If the lagrangian~(\ref{lagrangian}) is to yield $w_{\phi}<-1$ then (\ref{w})
implies
\begin{equation}
\label{gprimeconstraint}
g'(X)<0 \ ,
\end{equation}
where $g'(X)\equiv dg(X)/dX$ and we have used $f(\phi)\geq 0$ and $X\geq 0$. We
therefore require that $g(X)$ be a strictly monotonically decreasing function.
It is interesting to note in passing that, provided $f(\phi)$ is positive
semi-definite, the functional forms of $f(\phi)$ and $V(\phi)$ play no role in
determining whether $w_{\phi}$ is less than or greater than -1.
 
However, a constraint involving the potential does arise from the requirement
that the energy density of the theory satisfy $\rho>0$. This yields
\begin{equation}
\label{positiveenergy}
g(X)-2Xg'(X)<\frac{V(\phi)}{f(\phi)} \ .
\end{equation}
 
A necessary condition that the theory be stable is that the speed of sound of
$\phi$ be
positive~\cite{Garriga:1999vw} (see \cite{CHT} for a detailed stability
analysis of models
with $w_Q <-1$). This yields
\begin{equation}
\label{soundspeed}
c_s^2\equiv \frac{\partial p}{\partial \rho} =
\frac{p_{,X}}{\rho_{,X}}=\frac{g'(X)}{g'(X)+2Xg''(X)}>0 \ ,
\end{equation}
where a subscript $,X$ denotes a partial derivative with respect to $X$.
Since we have already specified $g'(X)<0$ this may be written as
\begin{equation}
\label{gdoubleprimeconstraint}
g''(X)<-\frac{g'(X)}{2X}
\end{equation}
 
Notice the difference between this class of models and the k-essence family in
terms of the potential $V(\phi)$ and the constraints placed on the functions
$g(X)$ and $f(\phi)$.
 
Let us illustrate these constraints with a simple example $g(X)=e^{-\alpha X}$,
with $\alpha>0$. This function trivially satisfies the
constraints~(\ref{gprimeconstraint}),(\ref{gdoubleprimeconstraint}). The
constraint~(\ref{positiveenergy}) then yields
\begin{equation}
\label{exampleconstraint}
\frac{V(\phi)}{f(\phi)}>(2\alpha X+1)e^{-\alpha X} \ .
\end{equation}
In the asymptotic regions this becomes $V/f>0$ as $X\rightarrow\infty$ and
$V/f>1$ as $X\rightarrow 0$. This may be satisfied without a particularly large
potential by arranging an appropriately behaved $f(\phi)$,
since~(\ref{positiveenergy}) constrains only the ratio $V/f$, making
fine-tuning issues less severe.
 
Let us now assume a (flat) Friedmann, Robertson-Walker (FRW) ansatz for the
space-time metric
\begin{equation}
\label{frw}
ds^2=-dt^2+a(t)^2 d\overrightarrow{x}^2 \ ,
\end{equation}
with $a(t)$ the scale factor. The resulting Einstein equations then become the
Friedmann equation
\begin{equation}
\label{friedmann}
H^2\equiv\left(\frac{{\dot a}}{a}\right)^2=\frac{8\pi G}{3}\rho
\end{equation}
and the acceleration equation
\begin{equation}
\label{acceleration}
\frac{{\ddot a}}{a}=-\frac{4\pi G}{3}(\rho +3p) \ .
\end{equation}
One then solves these equations along with those for the scalar field. In the
case of k-essence with $w_Q>-1$ it has been
shown~\cite{Armendariz-Picon:1999rj}-\cite{Armendariz-Picon:2000ah} that
{\it tracking} behavior can be obtained. This means that for a wide range of
initial conditions, the energy density of the field $\phi$ naturally evolves so
as to track the energy density in matter, providing some insight into why dark
energy domination began only recently in cosmic history. In our model, in
which we have included a potential for $\phi$ and are in the regime $w_Q<-1$,
the analysis becomes
somewhat more involved. Nevertheless, it can be shown that tracking behavior
persists.
However, as in all rolling scalar models, some fine-tuning remains, since one
must ensure the right amount of dark energy density today.
 
\section{Comparison with Observations: Method}
\label{method}
We restrict our analysis to flat models, for which
the effects of dark energy with $w_Q \geq -1$ on the angular power spectrum of
the CMB anisotropies have been carefully analyzed (see e.g.
\cite{rachel} and references therein).
The main effect of changing the value of $w_Q$ on the CMB anisotropies
is to introduce a shift by a linear factor ${\cal R}$ in the $l$-space
positions of the
acoustic peaks in the angular power spectrum~\cite{Bond:1997wr}. This shift is
given by
\begin{equation}
{\cal R}=\sqrt{(1-\Omega_Q)}y \ ,
\label{Req}
\end{equation}
where
\begin{equation}
y=\int_0^{z_{dec}}
\frac{dz}{\sqrt{(1-\Omega_Q)(1+z)^3+\Omega_{Q}(1+z)^{3(1+w_Q)}}} \ .
\end{equation}
 
\begin{figure}[thb]
\begin{center}
\includegraphics[scale=0.4]{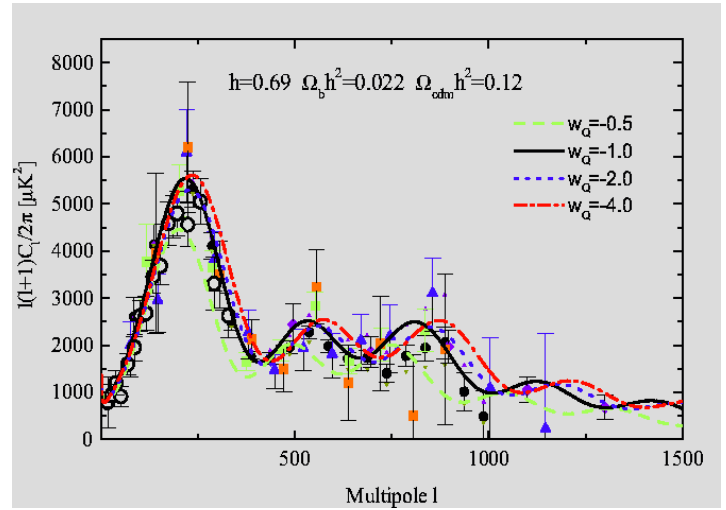}
\end{center}
\caption{The effect of varying $w_Q$ on the
COBE-normalized CMB angular power spectrum
and present CMB data. Since the shift of the power spectrum is
proportional to ${\cal R} \ell$,
the discrepancy is more important for higher values of $\ell$.}
\label{figomega}
\end{figure}

In order to illustrate this effect, we plot in Figure~\ref{figomega} a
set of theoretical power spectra, computed assuming
a standard cosmological model with the relative density in cold dark
matter $\Omega_{cdm}h^2=0.12$, that in baryons $\Omega_bh^2=0.022$,
with Hubble parameter $h=0.69$ but with $w_Q$ varied in the
range $(-4, -0.5)$. It is clear that decreasing $w_Q$ shifts the 
power spectrum towards smaller angular scales $\theta \sim l^{-1}$.
 
\begin{figure}[thb]
\begin{center}
\includegraphics[scale=0.4]{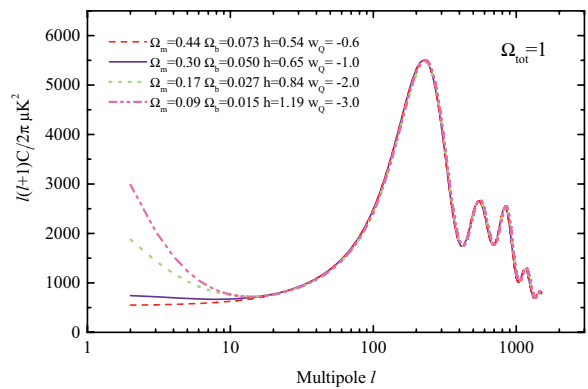}
\end{center}
\caption{Degenerate CMB power spectra
The models are computed assuming flatness
($\Omega_{tot}=\Omega_M+\Omega_Q=1$). On large angular scales the Integrated
Sachs Wolfe effect breaks the degeneracy for highly negative values of $w_Q$.
In general, the degeneracy of the spectra can be broken with a strong prior 
on $h$ or on $\Omega_M$.}
\label{figomega2}
\end{figure}

In considering the CMB power spectrum, it is
important to note that there is some degeneracy among the possible
choices of cosmological parameters.
First of all, the shift produced by a change in $w_Q$ can easily be 
compensated by a change in the curvature. However, degeneracies still
exist even when we restrict to flat models.
To emphasize this we plot
in Figure~\ref{figomega2} some degenerate spectra, obtained by keeping
$\Omega_bh^2$, $\Omega_Mh^2$, and ${\cal R}$ fixed, in a flat universe.
In practice, in order to preserve the shape of the spectrum
while decreasing $w_Q$, one has to increase $\Omega_Q$.
For flat models, one must therefore decrease $\Omega_M$ and,
since $\Omega_Mh^2$ must be constant, increase $h$.
Therefore, even if the CMB spectra are degenerate,
combining the CMB information with priors on $\Omega_M$ and
$h$ can be extremely helpful in bounding $w_Q$.
 
On large angular scales the time-varying Newtonian potential after decoupling
generates CMB
anisotropies through the Integrated Sachs-Wolfe (ISW) effect. This is clearly
seen in
Figure~\ref{figomega2} and is more pronounced for more negative values of
$w_Q$.
The effect depends not only on the value of
$w_Q$ but also on its variation with redshift. However, this is difficult to
disentangle from other cosmological effects.

\begin{figure}[thb]
\begin{center}
\includegraphics[angle=270,scale=0.3]{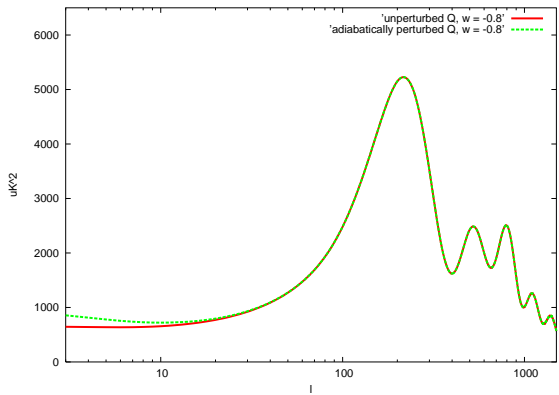}
\end{center}
\caption{Effect of including perturbations in the dark energy
fluid with a constant equation of state in the CMB power spectrum. 
Since dark energy dominates at late redshifts, the effect is present 
only on large angular scales.}
\label{perturbations}
\end{figure}

In all our analysis we will neglect perturbations in the dark energy
component. The reasons for this simplification are twofold:
on the one hand we prefer our analysis to remain as 
model-independent as possible, so that the results obtained here
are not affected by the choice of a particular dark energy model.
The study of the perturbations in particular quintessential
models goes beyond the scope of 
this paper. On the other hand, this approximation is also satisfied in a 
broad class of models and it is not completely straightforward 
to conclude that the inclusions of perturbations, while consistent 
with General Relativity, would yield a better approximation for a particular 
model of dark energy.
As an example, in Figure~\ref{perturbations}, we plot the CMB power
spectra for $w=-0.8$ computed with and without assuming
adiabatic perturbations in the dark energy fluid as in 
\cite{saralewis}. As we can see, since dark energy dominates 
the overall density  only well after recombination, the major effects 
are only on very large scales. We found that the inclusion of perturbations 
has no relevant effect on our results for models with $w>-1$ where
the computations of the perturbations are meaningful.

In order to bound $w_Q$, we consider a template of flat, adiabatic,
$Q$-CDM models computed with CMBFAST~\cite{sz}. We sample the
relevant parameters as follows:
$\Omega_{cdm}h^2 = 0.01,...0.40$, in steps of  $0.01$;
$\Omega_{b}h^2 = 0.001, ...,0.040$,
in steps of  $0.001$, $\Omega_{Q}=0.0, ..., 0.95$,
in steps of  $0.05$ and $w_Q=-3.0,...,-0.4$ in steps of $0.05$. 
Note that, once
we have fixed these parameters, the value of the Hubble constant is not an
independent parameter, since it is determined through the flatness condition.
We adopt the conservative top-hat bound $0.45 < h < 0.85$.

We allow for a reionization of the intergalactic medium by
varying the Compton optical depth parameter
$\tau_c$ over the range $\tau_c=0.05,...,0.40$ in steps of $0.05$.
 
For the CMB data we use the recent temperature and 
cross polarization results from the WMAP satellite
(\cite{Bennett:2003bz}) using the method explained in 
(\cite{Verde:2003ey}) and the publicly available code
on the LAMBDA web site.
We further include the
results from the BOOMERanG-98~\cite{ruhl},
DASI~\cite{halverson}, MAXIMA-1~\cite{lee},
CBI~\cite{pearson}, VSAE~\cite{grainge}, and Archeops~\cite{benoit}
experiments by using 
the publicly available correlation matrices and window functions.
We consider $7 \%$, $10 \%$, $4 \%$, $5 \%$, $3.5 \%$  and $5 \%$
Gaussian distributed
calibration errors for the Archeops, BOOMERanG-98, DASI, MAXIMA-1, VSA,
and CBI experiments respectively and include the beam uncertainties
using the analytical marginalization method presented in~\cite{bridle}.
The likelihood ${\cal L}$ for a given theoretical model is defined by
\begin{equation}
-2\ln{\cal L}=(C_B^{th}-C_B^{ex})M_{BB'}(C_{B'}^{th}-C_{B'}^{ex}) \ ,
\end{equation}
where  $M_{BB'}$ is the Gaussian curvature of the likelihood
matrix at the peak and $C_B$ is the theoreical/experimental signal 
in the bin (\cite{BJK}).
 
In addition to the CMB data we also consider
the real-space power spectrum
of galaxies in the 2dF $100$k galaxy redshift survey using the
data and window functions of the analysis of Tegmark et al.~\cite{thx}.
To compute the likelihood function ${\cal L}^{2dF}$ for the 2dF survey we
evaluate $p_i = P(k_i)$,
where $P(k)$ is the theoretical matter power spectrum
and $k_i$ are the $49$ k-values of the measurements in \cite{thx}.
Therefore
\begin{equation}
-2\ln{\cal L}^{2dF} = \sum_i \frac{[P_i-(Wp)_i]^2}{dP_i^2} \ ,
\end{equation}
where $P_i$ and $dP_i$ are the measurements and corresponding error bars
and $W$ is the reported $27 \times 49$ window matrix.
We restrict the analysis to a range of scales over which the fluctuations 
are assumed to be in the linear regime ($k < 0.1 h^{-1}\rm Mpc$).
When combining with the CMB data, we marginalize over a bias $b$
considered to be an additional free parameter.
 
We also incorporate constraints obtained
from the luminosity measurements of Type Ia supernovae (SN-Ia). In doing this
note that
the observed apparent bolometric luminosity $m_B$
is related to the luminosity distance $d_L$, measured in Mpc, by
$m_{B}=M+5 \log d_{L}(z)+25$,
where M is the absolute bolometric magnitude. Note also that
the luminosity distance is sensitive to the cosmological
evolution through an integral dependence on the Hubble factor
\begin{equation}
d_{L}=(1+z)\int_{0}^{z} dz'\frac{1}{H(z',\Omega_{Q},\Omega_{M},w_{q})} \ .
\end{equation}
We evaluate the likelihoods
assuming a constant equation of state, such that
\begin{equation}
H^2(z)=H_{0}^2\sum_{\alpha}\Omega_{\alpha}(1+z)^{(3+3w_{\alpha})} \ ,
\end{equation}
where the subscript $\alpha$ labels different components of the cosmological
energy budget.
The luminosity $m_{\rm eff}$ predicted from the observations is then calculated
by calibration
with low-z supernovae observations for which
the Hubble relation $d_{L}\approx H_{0}cz$ is obeyed.
We calculate the likelihood ${\cal L}^{\rm SN}$ using the relation
\begin{equation}
{\cal L}^{\rm SN}={\cal
L}_{0}\exp\left[-\frac{\chi^{2}(\Omega_Q,\Omega_M,w_Q)}{2}\right] \ ,
\end{equation}
where ${\cal L}_{0}$  is an arbitrary normalisation
and $\chi^{2}$ is evaluated using the observations of~\cite{super1}
and marginalising over $H_{0}$.
Finally, we also consider the $1\sigma$ contraint on the Hubble
parameter, $h=0.71\pm0.07$, obtained from Hubble Space Telescope
(HST) measurements~\cite{freedman}.

\section{Comparison with Observations: Results}
\label{results}
Table I shows the $2$-$\sigma$ constraints on $w_Q$ in a flat universe
for different
combinations of priors, obtained after marginalizing
over all remaining parameters.
 
\medskip
\begingroup\squeezetable
\begin{table}[bt]
\renewcommand*{\arraystretch}{1.5}
\label{tab}
\begin{tabular}{lrrrr}
\hline
\\
CMB+HST
&$-1.65<w_Q<-0.54$
\\
&$0.19<\Omega_M<0.43$
\\
CMB+HST+BBN
&$-1.61<w_Q<-0.57$
\\
&$0.20<\Omega_M<0.42$
\\
CMB+HST+SN-Ia
&$-1.45<w_Q<-0.74$
\\
&$0.21<\Omega_M<0.36$
\\
CMB+HST+SN-Ia+2dF
&$-1.38<w_Q<-0.82$
\\
&$0.22<\Omega_M<0.35$
\\
\hline
\\
\end{tabular}
\caption{
  Constraints on $w_Q$ and $\Omega_M=1-\Omega_Q$
  using different priors and datasets.
  We always assume flatness and that the age of the universe $t_0>10$~Gyr.
  The $2\sigma$ limits are found from the 2.5\%
  and 97.5\% integrals of the marginalized likelihood.
  The HST prior is $h=0.71 \pm0.07$, while for the BBN prior
  we use the conservative bound $\Omega_bh^2=0.020\pm0.005$.
}
\end{table}
\endgroup
 
It is clear that $w_Q$ is poorly constrained from CMB data alone,
even when the strong prior on the Hubble parameter from HST,
$h=0.71\pm0.07$, is assumed.
Adding a Big Bang Nucleosynthesis prior,
$\Omega_bh^2 =0.020 \pm 0.005$, has a small effect on the CMB+HST
result.
Adding SN-Ia data breaks the CMB $\Omega_M-w_Q$ degeneracy and
improves the limits on $w_Q$, yielding $-1.45< w_Q <-0.74$.
Finally, including data from the 2dF survey further breaks
the degeneracy, giving $-1.38 < w_Q <-0.82$ at $2$-$\sigma$.
Also reported in Table I are the constraints on $\Omega_M$.
The combined data suggests the presence of dark
energy with high significance, even if one only considers CMB+HST data.

\begin{figure}[t]
\begin{center}
\includegraphics[scale=0.4]{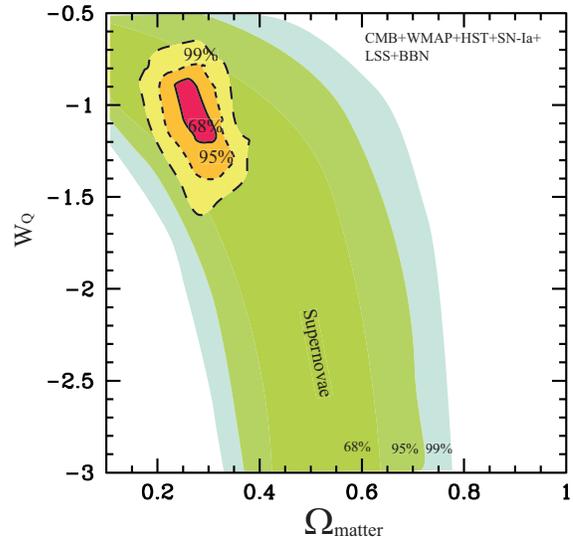}
\end{center}
\caption{Likelihood contours in the ($\Omega_M$, $w_Q$) plane for the
joint CMB+HST+SN-Ia+2dF analysis described in the text. We take the best-fit
values for the remaining parameters.
The contours correspond to 0.32, 0.05 and 0.01 of the peak value of the
likelihood, which are the 68\%, 95\% and 99\% confidence levels respectively.
Also plotted are the likelihood contours from type-Ia Supernovae alone.}
\label{figo1}
\end{figure}

It is interesting to project our likelihood onto the ($\Omega_M$, $w_Q$) 
plane. In Figure~\ref{figo1} we plot the likelihood contours in the
($\Omega_M$, $w_Q$) plane from our joint analyses of CMB+SN-Ia+HST+2dF
data. As we can see, there is strong supporting evidence for dark energy.
A cosmological constant with $w_Q=-1$ is in good agreement with all the
data and the most recent CMB results improve the constraints from 
previous and similar analyses (see e.g. \cite{mortsell}).

\section{Conclusions}
\label{conclusions}
In this paper we have provided new constraints on the dark energy
equation of state parameter $w_Q$ by combining recent
cosmological datasets.
We find $-1.38< w_Q <-0.82$ at the $95 \%$ confidence level, 
with best-fit model $\Omega_M=0.27$ and $w_Q=-1.05$.
A cosmological constant is a good fit to the data.
When comparison is possible (i.e. restricting to similar priors and
datasets), our analysis is compatible with other recent analyses
of $w_Q$ (e.g. see \cite{Spergel:2003cb}, \cite{rachel} and references
therein), however, our lower bound on $w_Q$ is
much tighter than the one recently reported in \cite{mortsell}.
In particular, we found that the CMB+HST dataset 
can already provide an interesting lower limit
on $w_Q$, while in \cite{mortsell} no constraint was obtained.
Part of the discrepancy can be explained by our updated CMB dataset
with the new Archeops, Boomerang, CBI, VSA and, mostly, WMAP results.
However, our CMB power spectra in Fig.1 are in disagreement with the
same spectra plotted in the Fig.2 of \cite{mortsell} where the
dependence on $w_Q$ seems limited only to the large-scale ISW term.

In the range $w>-1$ our results are in very good agreement with those
reported by Spergel et al. \cite{Spergel:2003cb}, which is using a 
different analysis method based on a Monte Carlo Markov
Chain and a slightly different CMB dataset. 

We found that including models with $w<-1$ does not significantly affect 
the results obtained under the assumption of $w>-1$.
In this respect, our findings are a useful complement to
those presented in \cite{Spergel:2003cb}.

As in \cite{Spergel:2003cb} and in most of previous similar analysis 
the constraints obtained here have been obtained under 
several assumptions: the equation of state is redshift independent, 
the perturbations in the dark energy fluid are negligible and/or
its sound speed never differs from unity in a significant way.
It is important to note that our result apply only to
models well described by these approximations. 

We have also demonstrated that, even by applying the most current constraints
on the dark energy equation of state parameter $w_Q$, there is much
uncertainty in its value. Interestingly, there is a distinct possibility
that it may lie in the theoretically under-explored region $w_Q <-1$. To
illustrate this we have provided a specific model in which $w_Q <-1$ 
is attained, and which satisfies the
assumption that $w_Q$ is approximately piecewise constant, as used in the data
analysis.
An observation of a component to the cosmic energy budget with $w_Q <-1$
would naturally have significant implications for fundamental physics.
Further, depending on the asymptotic evolution of $w_Q$, the fate of
the observable universe~\cite{Starkman:1999pg}-\cite{Huterer:2002wf} may be
dramatically altered, perhaps resulting in an instability of the
spacetime~\cite{CHT} or
a future singularity.
 
If we are to understand definitively whether dark energy is dynamical, and
if so, whether it is consistent with $w_Q$ less than or greater than $-1$,
we will need to bring the full array of cosmological techniques to bear on the
problem. An important contribution to this effort will be provided by direct
searches for supernovae at both intermediate and high redshifts~\cite{SNAP}.
Other, ground-based observations~\cite{LSST} will allow complementary analyses,
including weak gravitational lensing~\cite{Huterer:2001yu} and large scale
structure surveys~\cite{Hu:1998tk} to be performed.
 
At present, however, while the data remain
consistent with both a pure cosmological constant $\Lambda$, and with dynamic
classes of
models~\cite{Wetterich:fm}-\cite{Armendariz-Picon:2000ah},\cite{bastero-mersini}, 
nature may be telling us that the universe is an even stranger place than we
had imagined.
 
\medskip
 
\acknowledgments
 
We wish to thank Rachel Bean, Sean Carroll, 
Mark Hoffmann, Irit Maor, David Spergel, and Licia Verde 
for many helpful discussions,
comments and help. We acknowledge the use of CMBFAST~\cite{sz}.
The work of AM is supported by PPARC.
LM and MT are supported in part by the National Science
Foundation (NSF) under grant PHY-0094122 and LM is also supported in part by
the U.S. Department of Energy under contract number DE-FG02-85-ER40231.
CJO is supported by the Leenaards Foundation, the Acube Fund,
an Isaac Newton Studentship and a Girton College Scholarship.

\end{document}